# Mapping and Coding Design for Channel Coded Physical-layer Network Coding

Xu Li, Shengli Zhang, Gongbin Qian*

Department of Communication Engineering, College of Information Engineering, Shenzhen University

e-mail 07xli10@gmail.com, {zsl, qiangb}@szu.edu.cn

*Abstract*-**Although BICM can significantly improves the BER performance by iteration processing between the demapping and the decoding in a traditional receiver, its design and performance in PNC system has fewer studied. This paper investigates a bit interleaved coded modulation (BICM) scheme in a Gaussian two-way relay channel operated with physical layer network coding (PNC). In particular, we first present an iterative demapping and decoding framework specially designed for PNC. After that, we compare different constellation mapping schemes in this framework, with the convergence analysis by using the EXIT chart. It is found that the anti-Gray mapping outperforms the Gray mapping, which is the best mapping in the traditional decoding schemes. Finally, the numerical simulation shows the better performance of our framework and verifies the mapping design.**

*Index Terms*-**Physical layer network coding, iterative demapping and decoding, BICM, repeat accumulate code.**

## I. INTRODUCTION

The Physical-layer network coding (PNC) is a prospective technique in the wireless two-way relay communications [1], which exploits interference to boost the throughput rather than avoiding interference with orthogonal channels. Specifically, PNC including two transmission phases: uplink phase and the downlink phase. In the uplink phase, both end nodes transmit to the relay simultaneously and the relay directly transforms the received electromagnetic (EM) waves to the network coding form without detecting each individual packet. In the downlink phase, relay broadcasts the network coded signal to both end nodes. As shown in [2-3], PNC outperforms the traditional relay communication scheme and the straightforward network coding scheme significantly. In fact, PNC was further proved to be able to approach the capacity of the two-way relay channel (TWRC) in high SNR regions [4][5] and to approach the throughput upper bound of wireless networks [6].

To apply PNC in wireless systems, channel coding must be considered to combat the deterious effect of the wireless channels. The first systematic study of joint PNC and channel coding design was done in [7], where two baseline models were proposed to apply the channel coding: end-to-end and link-by-link. The former only takes the channel coding and decoding process in the end nodes, while the latter takes the process in both relay and end nodes [7]. In the first model, the relay just amplifies and forwards the additive EM waves without any decoding/encoding actions. The end nodes will extract the needed packets from the received signals, in which its own transmitted signal can be canceled beforehand. In the second model, the relay will first try to obtain the channel-decoded and network-encoded packets from the received additive EM waves, and then broadcast to the end nodes after channel encoding. Ref [7] focused on the link-by-link model for its better performance and investigated three different channel coding and decoding designs, including Chanel-decoding- Network-Coding (CNC) process Design1, Design2 (CNC2) and Arithmetic-sum CNC Design (ACNC). In [8], the CNC scheme with convolutional code was proposed. The ACNC scheme was then extended to the generalized form in [9]. In [10], the performance of ACNC scheme was analyzed with EXIT chart.

However, there are little work about the bit interleaved coded modulation (BICM) and mapping design in CNC, which is widely used in real wireless system due to the low complexity and good performance. By separating modulation and coding into two entities, Zehavi [11] recognized the performance of coded modulation over a Rayleigh fading channel can be further improved. Inspired by this discovery, BICM is organized by G. Caire [12] in 1996, whose basic structure is putting an bit-wise interleaver between the encoder and the modulator at the transmitter end. The iterative decoding of BICM (BICM-ID) can be traced back to [13], in which the BICM-ID significantly outperforms conventional BICM decoding scheme. Later in [14], several mappings of QAM have been investigated under the BICM-ID scheme, which demonstrated that BICM-ID with properly designed 16-QAM, can provide robust performance over both the AWGN and Rayleigh fading channel.

This paper focuses on the uplink phase and proposes a BICM encoding/decoding and PNC joint processing framework in TWRC. Within this framework, new algorithms are proposed and analyzed. In particular, the two end nodes adopt the same bit-wise interleaver and encoder. At the relay node, we propose an iterative channel decoding and network encoding algorithm. Similar to the point-to-point case, QPSK with anti-Gray mapping outperforms the Gray mapping.

This paper is organized as follows. In section II, we present the two-way relay system model adapted by this paper. In section III, we propose a BICM encoding and decoding framework for PNC. Under this framework, in section IV, we investigate the symbol mapping on the performance of the system by using the EXIT chart analysis [15]. Section V provides numerical results over Gaussian Channel. Section VI summarizes our works.

## II. SYSTEM MODEL

Consider a common two-way relay channel (TWRC) as shown in Fig.1, where two single-antenna end nodes $N_1$ and $N_2$ exchange information with the assistance of the relay node $N_3$. In our system, $S_i \in \{0,1\}^k$ denotes the length-$k$ uncoded binary packet of node $N_i$. After channel encoding, we get the length-$n$ encoded packet $X_i \in \{0,1\}^n$. In this paper, we use the quadrature phase-shift keying (QPSK) modulation, so the coded packet is mapped to the transmitted packet as $A_i \in \{1+i, 1-i, -1+i, -1-i\}^{n/2}$. The received noisy packet can be denoted as $Y_i = A_i + W_i$, in which $N_i$ is the Gaussian noise at the node $N_i$. In addition, we use the lower case letters, $s_i \in \{0,1\}$ , $x_i \in \{0,1\}$ , $a_i \in \{1+i, 1-i, -1+i, -1-i\}$ and

$y_i \in R$, to denote the source bit, channel encoded bit, QPSK modulated symbol and noise-polluted symbol in the corresponding packets, respectively. Note that the rate of the channel code is $R = k/n < 1$. We assume that each end node uses the same encoder in this paper.

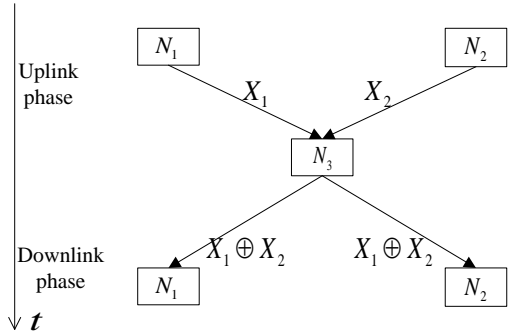

Fig. 1. Two-way relay channel transmission model

In this paper, we do not consider the direct link between the two end nodes due to the long distance. As a result, the communication between $N_1$ and $N_2$ can only be achieved through the relay node. What is more, considering the huge interference difference between transmitted signal and received signal at one node in real situation, all nodes are assumed to work in the half-duplex mode as in real wireless systems. Based on the two assumptions, at least two phases are needed to finish the information exchange between the two end nodes, which are the uplink phase and the downlink phase, as shown in Fig. 1. In the uplink phase, both the two end nodes transmit simultaneously to the relay node. Here, we also assume the two transmitted signals arrive at the relay node at a signal level synchronization[1]. Then we get the superimposed symbol at the relay node, which is expressed as:

$$y_3 = a_1 + a_2 + w_3 \quad (1)$$

where $w_3$ is the additive white Gaussian noise (AWGN) with variance $\sigma^2$, $a_i$ is the QPSK modulated symbol sent from node $N_i$. The task of relay node is to compute the network coded (NC) packet $S_3$ in the physical layer [1], i.e., try to generate $S_3 = S_1 \oplus S_2$ from the received packet $Y_3$.

In the downlink phase, the relay node will first channel encode $S_3$ into $X_3$ and then modulated it into $A_3$ before finally broadcasting it to both end nodes. Once the end nodes receive $A_3$, they can use the standard demapping and decoding process to obtain the network coded packet $S_3$. Then, the end nodes can recover the target information with the help of their self-information as $S_1 = S_2 \oplus S_3$ and $S_2 = S_1 \oplus S_3$.

As described above, the downlink processing in PNC is trivial while the uplink CNC processing, computing $S_3$ from $Y_3$ through channel decoding and PNC mapping [7], is the key. In this paper, we concentrate on the CNC process at the relay node with the QPSK mapping.

### III. BICM FRAMEWORK FOR PNC

This section shows more details about the encoding/decoding process dealing with PNC in TWRC system. We will first make a brief introduction of the framework in BICM encoder, which will be performed in both end nodes. After that, we elaborate the process at the relay node, which performs as an Iterative Demapping and decoding scheme.

#### A. Channel encoding model

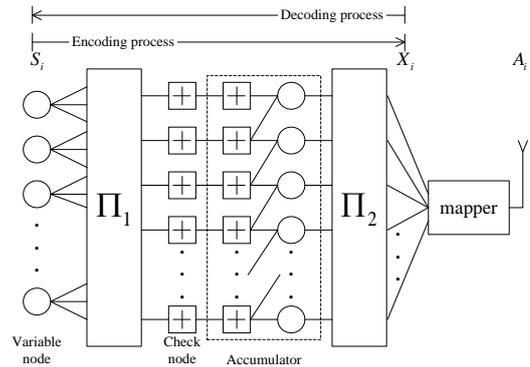

Fig. 2. Tanner graph of regular RA code with rate=1/3 ($d_v = 3, d_c = 1$), where the square elements denote the check node and the circular elements denote the variable in the Tanner graph. Note that, the encoding operation based on the Tanner Graph could be read from left to right, while the decoding operation is from right to left.

At the two end nodes, the repeat-accumulate (RA) code [16], as the Tanner graph depicted in Fig. 2, is used. Consider the encoding process by reading Fig. 2 from left to right. The source $N_i$ first produces a raw binary source packet $S_i$. After repetition process at the left-most variable node, the binary packet is interleaved by $\Pi_1$ and modulo-2 added at the check node, which is similar to the LDPC code[2]. The difference in the construction between RA code and LDPC code is the accumulator (ACC). We can decompose the ACC to be a concatenation of check nodes and variable nodes (VND) in the tanner graph, as shown in the dashed rectangle of Fig. 2. The ACC works as a differential encoder to create the connection between two adjacent check node. After the ACC, another bit-interleaver $\Pi_2$ helps to make the RA code more robust [17], which is actually indispensable in the BICM iterative decoding [13]. Then we get the encoded packet $X_i$, the QPSK mapper translates $X_i$ to the modulated packet $A_i$ before the transmission.

---

[1] Our paper focuses on the channel decoding. In fact, the asynchronized signal can be pre-transformed to the synchronized form before the decoding.

[2] As we only consider simple regular RA code with the degree-one check node (CND), the check node just pass the received message without modulo-2 addition.

## B. Iterative demapping and decoding

In this part, we try to explain the iterative process between demapping and decoding at the relay node. We will first introduce the general iterative framework by regarding the demapper and decoder as two separated modules. After that, we present the detailed implementation of the iteration process, which redesigns the inner iteration and out iteration to improve the performance.

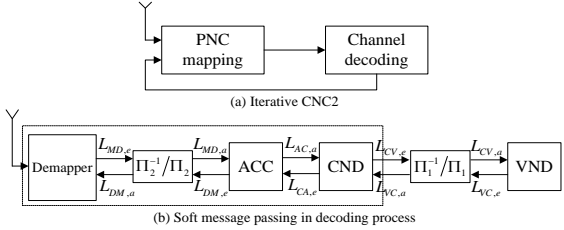

Fig. 3. Decoder model at the relay node

Different from the straightforward process of CNC2 [7], we propose an iterative demapping and decoding framework as in Fig. 3(a). The traditional CNC2 only takes the iteration in the decoding process while the scheme we proposed takes the iteration in both demapping and decoding process. In particular, the channel decoded soft symbols are feedback to the demapper to obtain a better demodulated symbol, which provides extrinsic information to improve the decoding performance.

We now elaborate the detailed iterative decoding processing as shown in Fig. 3(b). As in the figure, the decoding system is disassembled into four components: Demapper, ACC, CND and VND. The log-likelihood ratios (L-values or LLRs) of the extrinsic (*e*) message passed from the demapper (*M*) to the ACC (*A*) is denoted as $L_{MA,e}$. Similarly, $L_{AC,e}$ is from ACC to CND (*C*) and $L_{CV,e}$ is from CND to VND (*V*). Note that every extrinsic message is the a-priori message for the next component, so $L_{MA,e}$ is $L_{MA,a}$, $L_{CV,e}$ is $L_{CV,a}$.

### Demapping Processing:

We first introduce the demapper, which transforms the received signals to the soft message $L_{MA,e}$, i.e. the extrinsic message about the confidential of $X_1 \oplus X_2$. As required in the iterative CNC2, the demapper needs to calculate the L-value of every network coded bit by using the channel output $y_3$ and the feedback message $L_{AM,a}$. For each QPSK symbol $a_i$, the two bits modulated to it is denoted by $(x_{i,1}, x_{i,2})$. So, at the demapper, the L-value of bit $x_{3,1}$ conditioned on the received symbol $y_3$ can be calculated as

$$L(x_{3,1} | y_3) = \ln \frac{\Pr(x_{3,1} = 1 | y_3)}{\Pr(x_{3,1} = 0 | y_3)}$$
$$= \ln \frac{\Pr(x_{3,1}, x_{3,2} = 10 | y_3) + \Pr(x_{3,1}, x_{3,2} = 11 | y_3)}{\Pr(x_{3,1}, x_{3,2} = 00 | y_3) + \Pr(x_{3,1}, x_{3,2} = 01 | y_3)} \quad (2)$$

With the Bayes' rule $P(A|B) = P(B|A) \cdot P(A)/P(B)$, we get $\Pr(x_{3,1} x_{3,2} | y_3) = \Pr(y_3 | x_{3,1} x_{3,2}) \cdot \Pr(x_{3,1} x_{3,2})/\Pr(y_3)$. Due to the random bit-wise interleaver $\Pi_2$, the two bits $x_{3,1}$ and $x_{3,2}$ can be regarded as independent with each other. Then we can rewrite (2) as

$$L(x_{3,1} | y_3, L_{AM,a})$$
$$= L_{AM,a}(x_{3,1}) + \ln \frac{\Pr(y_3 | x_{3,1} x_{3,2} = 10) + \Pr(y_3 | x_{3,1} x_{3,2} = 11) \cdot \exp[L_{AM,a}(x_{3,2})]}{\Pr(y_3 | x_{3,1} x_{3,2} = 00) + \Pr(y_3 | x_{3,1} x_{3,2} = 01) \cdot \exp[L_{AM,a}(x_{3,2})]} \quad (3)$$

where $L_{AM,a}(x_{3,2}) = \ln \frac{\Pr(x_{3,2} = 1)}{\Pr(x_{3,2} = 0)}$ is the a-priori message of $x_{3,2}$, and its initial value is zero, which means there is no a-priori knowledge about the coded bits yet. Finally, the extrinsic information can be obtained by subtracting the a-priori information as

$$L_{MA,e}(x_{3,1} | y_3, L_{AM,a}) = L(x_{3,1} | y_3, L_{AM,a}) - L_{AM,a}(x_{3,1})$$
$$= \ln \frac{\Pr(y_3 | x_{3,1} x_{3,2} = 10) + \Pr(y_3 | x_{3,1} x_{3,2} = 11) \cdot \exp[L_{AM,a}(x_{3,2})]}{\Pr(y_3 | x_{3,1} x_{3,2} = 00) + \Pr(y_3 | x_{3,1} x_{3,2} = 01) \cdot \exp[L_{AM,a}(x_{3,2})]} \quad (4)$$

### Decoding Processing:

We then introduce the signal processing in the decoder. After de-interleaving $L_{MA,e}$, the channel decoder performs the traditional RA iterative decoding process by passing the soft message through ACC decoder, CND and the VND in a successive way. In the Tanner graph of Fig. 2, the output extrinsic message associated to an output edge-$i$ on a check node $p$ can be expressed as

$$L_{i,e}(x_p) = \ln \frac{\Pr(x_p = 1)}{\Pr(x_p = 0)} = \ln \frac{1 - \prod_{j \neq i} \frac{1 - \exp^{L_{j,in}}}{1 + \exp^{L_{j,in}}}}{1 + \prod_{j \neq i} \frac{1 - \exp^{L_{j,in}}}{1 + \exp^{L_{j,in}}}} \quad (5)$$

where $L_{j,in}$ represents the incoming message associated to edge-$j$. For a variable node $q$, the output extrinsic message of edge-$i$ can be computed as (4).

$$L_{i,e}(x_q) = \ln \frac{\Pr(x_q = 1)}{\Pr(x_q = 0)} = \sum_{j \neq i} L_{j,in} \quad (6)$$

### Inner iteration and out iteration:

The iteration between demapper and the decoder is shown in Fig. 3(b). The demapper, the ACC and the CND forms a single unit. The iterative processing is first performed within this unit and it is referred to as the inner iteration. The purpose of the inner iteration is to adequately infuse the a-priori knowledge $L_{VC,a}$ [18]. The out iteration will then be taken between the Inner iteration and the VND. In this paper, three inner iteration is performed in each out iteration.

## IV. MAPPING DESIGN BASED ON EXIT CHART

In traditional decoding, the demapping and decoding processes are separated. Then the Gray mapping performs best by minimizing the detection BER. However, in BICM receiver, non-Gray mapping may performs better by providing extrinsic information in the iterations. Therefore, mapping design is an important problem in BICM systems. When PNC is considered, the mapping design is more complicated since the constellation at the relay node is a superimposed one of the two constellations at the transmitters. This section will use the EXIT chart to analysis the mapping design on the performance of the proposed BICM scheme.

As proved in [19] [20], the symbol-wise mutual information $I(a_3; y_3)$ is independent of the mapping mechanism. With the chain rule of mutual information [21], the mapping only influences the partitioning of the total amount of bit-wise mutual information $I(x_3; y_3)$. Although the mapping scheme cannot increase the total mutual information, the selection of mapping scheme can affect the performance of decoding algorithm with iteration. We focus on the impacts of different mapping schemes in our proposed iterative CNC2.

We consider QPSK mapping in iterative CNC2 model. Each QPSK mapping scheme shown in Fig. 4 contains a traditional QPSK mapping and the corresponding PNC QPSK mapping. Note that the hollow mapping is applied at the end nodes and solid mapping is applied at the relay node. We can easily find that the PNC QPSK mapping is many point to one sequence mapping (when detection), which is unlike the traditional constellation. For example, the binary sequence $x_{3,1}x_{3,2} = 00$ could correspond to four constellation points in Fig. 4. Because of this special character, the PDF of the $y_3$ will be a sum form as

$$\Pr(y_3 | x_{3,1}x_{3,2}) = \sum_{a_i \equiv (x_{3,1}x_{3,2})} \Pr(a_i) \cdot \Pr(y_3 | a_i)$$

$$= \sum_{a_i \equiv (x_{3,1}x_{3,2})} \Pr(a_i) \cdot \frac{1}{2\pi\sigma^2} \exp\left(-\frac{1}{2\sigma^2}\|y_3 - a_i\|^2\right) \quad (7)$$

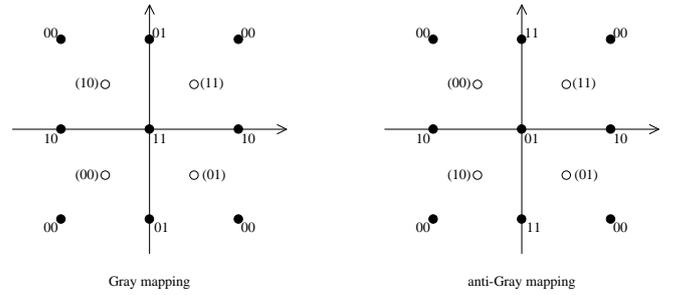

Fig. 4. Superimposed Constellations of Two QPSK Signals in CNC2, the hollow constellation points denote the QPSK modulation of one node and solid ones denote the superimposed QPSK modulation. Every solid constellation point is the physical addiction of two hollow constellation points.

There are two different mappings for QPSK modulation, Gray mapping (GM) and anti-Gray mapping (AGM). One visual difference between such two mapping schemes is the hamming distance distribution, either the traditional QPSK mapping or the PNC mapping, as shown in Fig. 4. Similar to the work done in [20], we draw the EXIT charts of the two mapping schemes, as given in Fig. 5, where $I_A$ is the prior input to the demapper, $I_E$ is the extrinsic output of the demapper. Obviously, the prior message makes little sense to the demapper in Gray mapping, which means when considering the iterative framework with Gray mapping, the performance nearly has no improvement for iterations. But when the anti-Gray mapping is employed, the iterative framework could utilize the feedback message to get a better output for the subsequent channel decoder. The $I_E$ of Gray mapping is higher than that of anti-Gray mapping before the cross point, and after the cross point the former is lower than the latter. That means the anti-Gray mapping can only outperforms the Gray mapping at the higher $I_A$, i.e., only when the feedback message is good enough. Therefore, as the iteration goes, the anti-Gray mapping scheme works better than the Gray mapping in our iterative framework.

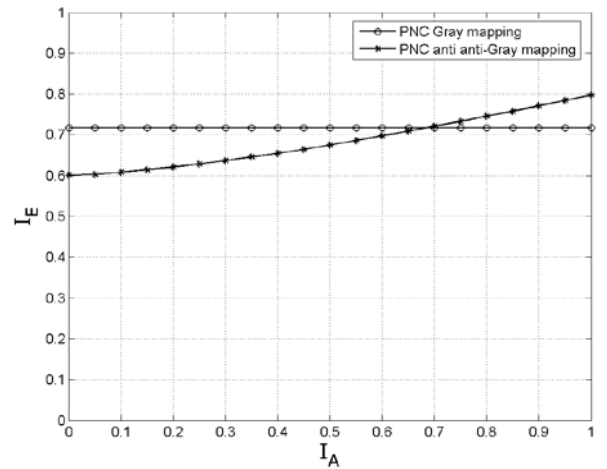

Fig. 5. EXIT charts of two demappers at SNR=4dB

When considering the whole decoding system at the relay node, i.e., concatenating the demapper and decoder, we can draw the EXIT chart to reveal the convergence character of the decoding scheme. For contrast, we draw the EXIT curves for the inner iteration with different mapping schemes and the VND. Fig. 6 shows us three EXIT curves of a regular RA code with $R=1/3$ at $E_b/N_0 = 1.8dB$, the upper two are the EXIT curves of inner iteration, the lowest one is for the VND. We can easily find that in such case, the curve of both AGM and GM stays above the curve of VND, which indicates that either mapping scheme could correctly recover the network coded data from the received signal. We also notice that before the cross point the GM works better than the AGM, but as the iteration goes, AGM outperforms GM after the cross point.

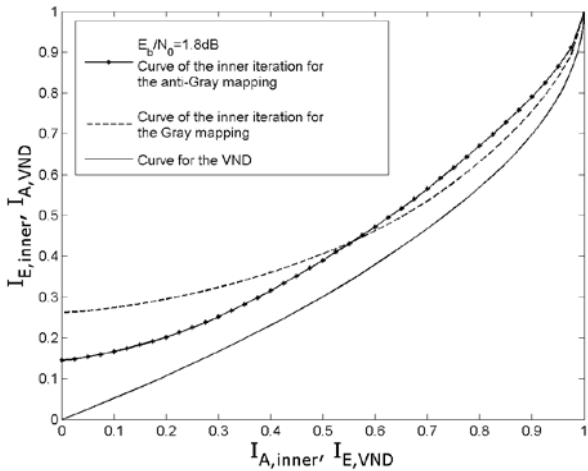

Fig. 6. EXIT curves of AGM and GM for an regular RA code with $R=1/3$ at $E_b/N_0 = 1.8dB$

## V. NUMERICAL SIMULATION

In this section, we investigate the BER performance of the above iterative decoding scheme. We employ the regular RA code with $R=1/3$ in our simulations. The noise is AWGN with variance $\sigma^2$ and the SNR is defined as $2/\sigma^2$ (the average power of one end node is 2 as we use the QPSK modulation). We calculate the BER with the decoded packet $S_3$ and the network coded genetic packet $S_1 \oplus S_2$.

In Fig. 7, we show the BER performance of iterative CNC2 and the traditional CNC2 under the two mapping schemes that we consider above. In the simulation, the uncoded packet length is set to 4096 and the BER is calculated over 1000 packets, with 20 iterations in the decoding algorithm. The result shows without the iteration between demapper and channel decoder, the anti-Gray mapping scheme performs badly, but the iteration included the demapper nearly has no effect to the Gray mapping scheme. That means the demapper iteration is very important to the anti-Gray mapping, but not to the Gray mapping. The anti-Gray mapping can outperform the Gray mapping quickly as the SNR increase.

In Fig. 8, we investigate the impact of packet length (2048 and 4096 bits) to the BER performance of the iterative CNC2 decoding scheme with anti-Gray mapping and Gray mapping under different SNR. The result shows that the longer packet length is, the better performance it will get.

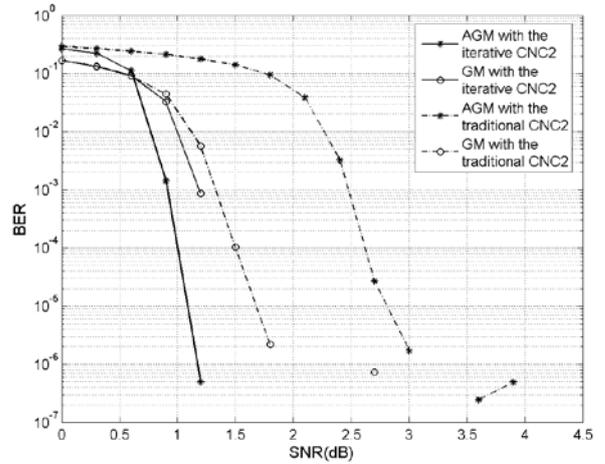

Fig. 7. BER performance of iterative CNC2 and traditional CNC2 under different mapping schemes.

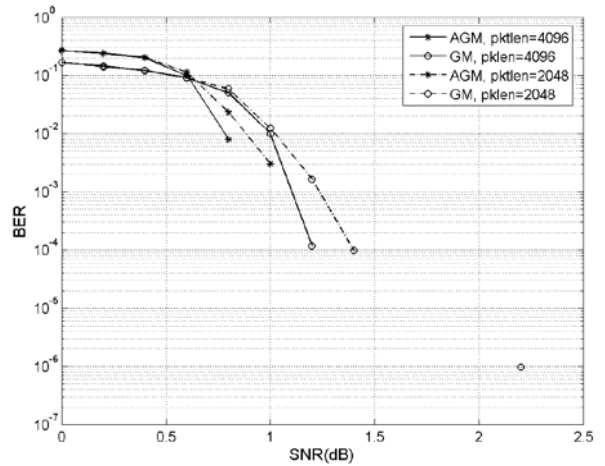

Fig. 8. BER performance of AGM and GM with different packet length.

## VI. CONCLUSION

In this paper, we have proposed an iterative demapping and decoding framework at the relay node in TWRC with PNC. The EXIT chart is used to study the convergence behavior of the decoding scheme. We found that under such decoding system, the anti-Gray mapping scheme outperforms the Gray mapping significantly.

ACKNOWLEDGEMENT

This paper is partially supported by NSFC (No. 61372078, 60602066, and 60773203) and supported by Foundation of Shenzhen City (No. JC201005250034A , JCYJ20120613174214967,